\documentclass{ws-procs9x6}

\begin{document}

\title{UNITARITY, GHOSTS AND NONMINIMAL TERMS IN LORENTZ VIOLATING QED}

\author{CARLOS M. REYES }

\address{ Departamento de Ciencias 
B{\'a}sicas, Universidad del B{\'i}o B{\'i}o,\\ 
Chillan, Casilla 447, Chile \\
E-mail: creyes@ubiobio.cl}
\begin{abstract}
The unitarity of a Lorentz-invariance violating QED model with 
higher-order Myers and Pospelov photons coupled to standard fermions is studied.
As expected, we find ghost states associated to the higher-order terms that may lead to the loss of unitarity.
An explicit calculation to check perturbative unitarity 
in the process of electron-positron scattering   
is performed and it is found to be possible to be preserved. 
\end{abstract}
\bodymatter
\section{Introduction}
In recent years, higher-order operators have become the object of 
intense study in the search for possible effects of Lorentz invariance 
violation \cite{M-P,nSME,rad-corrections}. These Planck-mass suppressed higher-order operators 
allows to describe new physics beyond those obtainable from renormalizable operators, 
that is, operators with mass dimension four or less \cite{Colladay-Kostelecky,Data}.
For example, the higher-order effective theory 
may involve additional degrees of freedom associated to ultra-high energies which do not converge perturbatively
to the normal ones when taking the limit of
the dimensionless parameters in the effective terms to zero. Lee and Wick studied these exotic modes in the context of
negative metric theories \cite{Lee-Wick} and in spite of the 
ghost states that appear, they showed that unitarity can be preserved by
demanding all stable particles to be positive norm states \cite{Lee-Wick,CUTW}.

Here we check perturbative unitarity in a QED 
consisting of
higher-order Myers and Pospelov photons \cite{M-P} 
and standard fermions.
\section{The Myers and Pospelov model}
The Myers-Pospelov Lagrangian density for photons is given by
\begin{eqnarray}\label{M-M-P}
\mathcal L=-\frac{1}{4}F^{\mu\nu}  F_{\mu\nu}- \frac{\xi}{2M_P} 
n_{\mu}\epsilon^{\mu\nu \lambda \sigma} A_{\nu}(n 
\cdot \partial)^2   F_{\lambda \sigma},
\end{eqnarray}
where $n$ is a four-vector defining a preferred 
reference frame, $M_P$ is the Planck mass and $\xi$ is a dimensionless parameter. 

We can always select a real basis
of four-vectors $e_{\mu}^{(a)}$ to be orthonormal 
and to satisfy the properties described in Ref.\ \refcite{unitarityQED}.
In analogy with the left and right handed polarizations of usual electrodynamics 
we can switch to a basis of complex four-vectors $\varepsilon^{\lambda}_{\mu}$ and define 
the orthogonal projectors $P^{\lambda}_{\mu \nu}$ as
\begin{eqnarray}\label{projector} 
\varepsilon^{\lambda}_{\mu}
 =\frac{1}{\sqrt{2}}(e_{\mu}^{(1)}+i\lambda \,e_{\mu}^{(2)}),
\qquad 
P^{\lambda}_{\mu \nu}= - \varepsilon^{\lambda}_{\mu}
\varepsilon^{*\lambda}_{\nu},
\end{eqnarray}
where $\lambda =\pm$.
To derive the dispersion relation we can expand
the gauge field in term of this complex basis and replace in
 the equations of motion
to arrive at 
\begin{eqnarray}
(k^2)^2-4g^2(n \cdot k)^4 \left((n\cdot k)^2-n^2k^2\right)=0,
\end{eqnarray}
in agreement with the work in Ref.\ \refcite{formal}.
\section{Unitarity}
Here we check perturbative unitarity in the process of electron-positron scattering
$e^+e^-\to e^+e^-$. For this we use the optical theorem 
which relates the imaginary part of the forward scattering 
amplitude $\mathcal M_{ii}$ with the total cross section as
\begin{eqnarray}\label{unitarity}
2 \,{\rm {Im}} \mathcal M_{ii}= \sum_{m} \int d\Pi_m
\left| \mathcal M_{im} \right|^2,
\end{eqnarray}
where the sum runs over all intermediate physical states.

Considering the QED extension model 
the amplitudes that contribute to the $S$-matrix are the direct amplitude
\begin{eqnarray}\label{ampdir}
\mathcal M^{{\rm dir }}=  (-ie)^2 \int d^4 k\, \delta^{4} 
(p_1-p_1^{\prime}-k) \widehat U^{\mu} U^{\nu} G_{ \mu \nu }(k),
\end{eqnarray}
and the exchange amplitude
\begin{eqnarray}\label{amplitud2}
\mathcal M^{{\rm ex }}=  (-ie)^2 \int d^4 k\, \delta^{4}
 (p_1+p_2-k) \widehat V^{ \mu} V^{\nu} G_{ \mu \nu }(k),
\end{eqnarray}
where $\widehat U^{ \mu}=  N_{p_2} N_{p_2^{\prime}} \bar v(p_2) 
\gamma^{\mu}v(p_2^{\prime})$, $U^{\nu}= N_{p_1^{\prime}} N_{p_1} 
\bar u(p_1^{\prime}) \gamma^{\nu}u(p_1)$ and 
$\widehat V^{ \mu}= N_{p_1^{\prime}} N_{p_2^{\prime}} \bar
 u(p_1^{\prime}) \gamma^{\mu}v(p^{\prime}_2)$, $V^{\nu}=
 N_{p_2} N_{p_1}
\bar v(p_2) \gamma^{\nu}u(p_1)$ and where $N_{p}=
\sqrt{ \frac{m}{E_{p}}}$ is the usual fermionic normalization constant.

Let us start with the left hand side of the unitarity condition (\ref{unitarity}).
A similar calculation has been given in
 the minimal sector of the Standard-Model Extension, see Ref.\ \refcite{Marco-Schreck}.
To simplify we will consider 
the lightlike case where we have a ghost state with frequencies 
$\omega_0^{\lambda}$ and two photons with frequencies 
$\omega_{1,2}^{\lambda}$, see Ref.\ \refcite{unitarityQED}, and the propagator
\begin{eqnarray}\label{PROPAGATOR}
G_{ \mu \nu }(k)&=&-\sum_{\lambda} 
\frac{P^{\lambda}_{\mu \nu}(k)}{ k^2+2g\lambda(n\cdot k)^3+
i\epsilon  },
\end{eqnarray}
where and we have included the $i\epsilon$ prescription. 

We are interested in the imaginary part of the 
forward scattering amplitude, therefore let us set
$p_1^{\prime}\to p_1$ and $p_2^{\prime}\to p_2$. 
Moreover, we can see that the direct process does not 
contribute since the virtual photon can never be on shell
for non-zero external momenta, hence ${\rm Im}[\mathcal M^{{\rm dir }}]=0$.
Let us find the contribution of the exchange process 
and substitute the propagator (\ref{PROPAGATOR}) in (\ref{amplitud2})
\begin{eqnarray}
\mathcal M^{{\rm ex }}=e^2  
\int \frac{d^4 \vec k}{(2\pi)^4}\delta^4(p_1+p_2-k)  
V^{ \mu} V^{* \nu}
\sum_{\lambda} \frac{P^{\lambda}_{\mu \nu}(k)}{ 
 k^2+2g\lambda(n\cdot k)^3+i\epsilon  }.
\end{eqnarray}
 Because only the poles can contribute to the
 imaginary part and due to energy conservation encoded 
in $\delta^4(p_1+p_2-k)$, we have that only the positive poles of the virtual 
photon have a chance to contribute. 
We can discard the ghost contribution since its
 energy $  | \omega_0^{\lambda}| \sim 1/2g$ lies 
beyond the region of validity of the effective theory. That is, the external fermions
will always satisfy the condition $p_{01}+p_{02}<|\omega_0^{\lambda}|$. Hence, we have
\begin{eqnarray}
&&2{\rm Im}[\mathcal M^{{\rm ex }}]\nonumber \\&=& - e^2  \int d k^0
\int \frac{d^3 \vec k}{(2\pi)^3}\delta^4(p_1+p_2-k)        
 V^{\mu} V^{* \nu} \sum_{\lambda}\frac{  
 P^{\lambda}_{\mu \nu}  \delta (k_0-\omega_1^{\lambda})}
{2g\lambda (k_0-\omega_0^{\lambda}) (k_0-
\omega_2^{\lambda}) }, \nonumber
\\ 
&=& e^2\int \frac{d^3 k}{(2\pi)^3   }\delta^4(p_1+p_2-k)  
   V^{\mu}   V^{* \nu} \sum_{\lambda} \frac{
 \varepsilon_{\mu}^{\lambda}  \varepsilon_{\nu}^{*\lambda} }{2g\lambda 
 (\omega_1^{\lambda}-\omega_0^{\lambda})
 (\omega_1^{\lambda}-\omega_2^{\lambda})},  \nonumber \\
&=& \int \frac{d^3 k}{(2\pi)^3 }\delta^2(p_1+p_2-k) 
 \sum_{\lambda}
  \left| \mathcal  M_{\lambda} \right|^2,
\end{eqnarray}
where we have used the notation $\mathcal 
 M_{\lambda} =(-ie) N_{k,\lambda} V^{\mu}  
\varepsilon_{\mu}^{\lambda}  $ 
 for the physical process
$\mathcal M_{\rm phys}(e^+e^-\to \gamma)$
and we have introduced the normalization 
constant $N_{k,\lambda}=\frac{1}{\sqrt{2g\lambda 
 (\omega_1^{\lambda}-\omega_0^{\lambda})
 (\omega_1^{\lambda}-\omega_2^{\lambda})}} $.
Finally we have
\begin{eqnarray}
2{\rm Im}[\mathcal M]= \int \frac{d^3 k}{(2\pi)^3 }
\delta^2(p_1+p_2-k) 
  \left| \mathcal  M_{\rm phys} \right|^2,
\end{eqnarray}
and therefore the unitarity condition is satisfied
 in this scattering process.
\section{Conclusions}
With an explicit calculation we have verified that 
the unitarity condition 
in the process of electron-positron scattering at
tree level order is satisfied. 

A next step is to verify unitarity to order $e^2$ that will require to analyze 
more diagrams. Some of them contain loops 
where the ghosts can appear off-shell, thus, introducing an extra difficulty. 
Checking the unitarity condition 
to these order will give us a robust support in order to make physical predictions
in the theory.
\section*{Acknowledgments}
This work was supported by the Direcci\'on de Investigaci\'on de
la Universidad del B\'{\i}o-B\'{\i}o grant 123809 3/R and FAPEI.

\end{document}